\documentclass[superscriptaddress,longbibliography,aps,prl,reprint,showpaFcs]{revtex4-1}

\pdfoutput = 1

\usepackage[T1]{fontenc}  
\usepackage[ngerman, english]{babel}

\usepackage{textcomp}  

\usepackage{graphicx,color}
\usepackage{graphicx}  
\usepackage{subfigure} 
\usepackage[export]{adjustbox}

\usepackage{siunitx}  
\usepackage{amsmath}

\begin{document}

\title{Signatures of strong coupling on nanoparticles: Revealing absorption anticrossing by tuning the dielectric environment}
\author{F. Stete}
\affiliation{Institut f\"ur Physik \& Astronomie,
  Universit\"at Potsdam, Karl-Liebknecht-Str. 24-25, 14476 Potsdam,
  Germany}
 \affiliation{Humboldt-Universit\"at zu Berlin, School of Analytical Sciences Adlershof (SALSA), Unter den Linden 6, 10999 Berlin, Germany}
\author{ W.Koopman}
\affiliation{Institut f\"ur Physik \& Astronomie,
  Universit\"at Potsdam, Karl-Liebknecht-Str. 24-25, 14476 Potsdam,
  Germany}
 \author{ M.Bargheer}
 \affiliation{Institut f\"ur Physik \& Astronomie,
  Universit\"at Potsdam, Karl-Liebknecht-Str. 24-25, 14476 Potsdam,
  Germany}
\affiliation{Helmholtz Zentrum Berlin, Albert-Einstein-Str. 15, 12489
  Berlin, Germany}

\begin{abstract}
Strongly coupled plasmon-exciton systems offer promising applications in nanooptics. The classification of the coupling regime is currently debated both from experimental and theoretical perspectives. We present a method to unambiguously identify strong coupling in plasmon-exciton core-shell nanoparticles by measuring true absorption spectra of the system. We investigate the coupling of excitons in J-aggregates to the localized surface plasmon polaritons on gold nanospheres and nanorods by fine-tuning the plasmon resonance via layer-by-layer deposition of polyelectrolytes. While both structures show a characteristic anticrossing in extinction and scattering experiments, the careful assessment of the systems' light absorption reveals that strong coupling of the plasmon to the exciton is only present in the nanorod system. In a phenomenological model of two classical coupled oscillators, intermediate coupling strengths split up only the resonance frequency of the light-driven oscillator, while the other one still dissipates energy at its original frequency. Only in the strong-coupling limit, both oscillators split up the frequencies at which they dissipate energy, qualitatively explaining our experimental finding.
\end{abstract}

\maketitle

The electromagnetic coupling of molecular excitations to plasmonic nanoparticles offers a promising method to manipulate the light-matter interaction at the nanoscale. This approach is frequently used to enhance the optical cross-section of molecules, e.g. in surface enhanced Raman scattering (SERS) \cite{Stiles.2008}, enhancement of fluorescence \cite{Kinkhabwala.2009} and infrared absorption \cite{Neubrech.2008, Adato.2009}, plasmon-enhanced light- harvesting in dye-sensitized solar cells \cite{Qi.2011} or plasmon-enhanced dye-lasers \cite{Noginov.2009}.

The coupling strength between molecular excitations and plasmons is given by the rate of energy-exchange between the two components $\Omega = \textbf{\textit{E}} \cdot \boldsymbol{\mu}/\hbar$ \cite{Vasa.2013}. Here $\boldsymbol{\mu}$ describes the transition dipole moment of the emitter and $\textbf{\textit{E}}$ the electric-field strength of the light at the emitter-location. Plasmonic nanoparticles foster exceptionally high coupling strengths, due to their capacity to strongly concentrate the light-field to sub-wavelength mode volumes and hence to generate very high electrical field-strengths in their vicinity. A particularly interesting coupling regime occurs, if the coupling increases to a level such that $\Omega$ surpasses all damping rates in the system. In this so-called strong coupling regime hybrid light-matter states emerge, which cannot be divided into separate light and matter components. The new resonances of the coupled system emerge from the plasmon resonance $\omega_p$ and the exciton resonance $\omega_{ex}$ as \cite{Zengin.2015, Rudin.1999}

\begin{equation}
\omega_{\pm} = \frac{1}{2}(\omega_{p}+\omega_{ex}) \pm \sqrt{\frac{1}{4}(\omega_{p}-\omega_{ex})^2+g^2} ,
\label{anticrossingequation}
\end{equation}

where g is the coupling parameter.

Hence, the presence of the new hybrid-states can be detected by observing characteristic optical spectra showing two new resonances with a separation of $\Delta = (\omega_+ - \omega_-)$. 

Realizing strong light-matter coupling on the nanoscale promises both, interesting possibilities for the fundamental study of light-matter interaction as well as a great potential for applications. Suggestions include: threshold-less lasing \cite{Torma.2015}, all optical switching \cite{Vasa.2010}, the manipulation of chemical reactions \cite{Hutchison.2012}, the adjustment of work-functions \cite{Hutchison.2013} and in particular applications in the context of nanoscale quantum optics \cite{Torma.2015} such as quantum encryption and optical quantum computing.

The most frequently used approach for achieving strong light-matter coupling on nanoparticles, is the fabrication of hybrid core-shell particles with a noble-metal core and a molecular shell. Several hybrid nanoparticles that allegedly show strong coupling have been presented in recent literature \cite{Fofang.2008, DjoumessiLekeufack.2010, Uwada.2007, Balci.2014,Ni.2008}. In several cases these claims have however been challenged \cite{Zengin.2016, Antosiewicz.2014}, as an unambiguous determination of the coupling regime for hybrid nanoparticles is difficult. A clear prove for the presence of strong coupling would be obtained by comparing $\Omega$ to the spectral line-width of both the uncoupled plasmon and the molecules. For nanoparticles however, the spectral line-widths are frequently masked by inhomogeneous broadening, mostly due to particle-size dispersion and local variations in the chemical environment. The same effects cause different particles to feature different coupling-strengths and consequently different spectral splittings, which may result in an underestimation of the real coupling strengths when measuring the ensemble spectrum.

Due to the difficulties in determining the correct linewidths and splitting from the ensemble spectrum, the presence of strong coupling is often claimed on the basis of observing a "dip" in the extinction \cite{Fofang.2008, Fofang.2011, DjoumessiLekeufack.2010, Melnikau.2016} or scattering spectrum \cite{Uwada.2007}, which is interpreted as a spectral splitting into new states. An anticrossing in the functional dependence of the observed resonance frequencies on the uncoupled plasmon frequency $\omega_p$ \cite{Cade.2009} frequently serves as further evidence. The incautious use of extinction or scattering spectra for discussing the coupling strength of hybrid particles can however easily lead to erroneous conclusions \cite{Zengin.2016, Antosiewicz.2014}. In particular, real strong coupling can be confused with effects like Fano resonances or induced transparency \cite{Antosiewicz.2014, Faucheaux.2014}, which rather represent an intermediate coupling regime. This brings up the question, how to prove the presence of strong coupling for an ensemble of hybrid nanoparticles. Several authors discussed that absorption and fluorescence are the only reliable quantities to determine the presence of strong light-matter coupling \cite{Antosiewicz.2014, Zengin.2016}. However, both quantities are not regularly determined for nanoparticle ensembles.

This letter presents an experimental approach to determine the coupling regime of hybrid nanoparticles that successfully distinguishes the intermediate and strong coupling regimes. It describes an experimental procedure: a) to measure the real absorption spectrum of particle ensembles, and b) to determine the anticrossing relation without nanoparticle size variation, but via shifting the plasmon resonance by layer-by-layer deposition of polyelectrolytes \cite{Mitzscherling.2015}. The different coupling regimes are illustrated by two core-shell nanoparticle systems, gold nanorods and nanospeheres, which are both coated with an excitonic molecular shell. Both particle types have been claimed to support strong coupling \cite{Uwada.2007, Melnikau.2016} on grounds of an observed dip in the extinction spectrum. However, while for rods the presence of strong coupling has been confirmed by fluorescence measurements \cite{Melnikau.2016}, theory shows that gold nanospheres cannot reach the strong coupling regime for fundamental reasons \cite{Faucheaux.2014}. The absorption measurements discussed in this letter support this classification. Using a classical coupled pendulum model with an appropriate coupling to the periodic excitation, we discuss which information can be obtained from scattering and absorption measurements and illustrate the underlying mechanism leading to the observed differences in absorption, extinction and scattering spectra.

The samples used in the experiments were based on $\SI{100}{nm}$ gold-nanospheres and $\SI{25}{nm} \times \SI{56}{nm}$ gold nanorods, coated with a shell of \textit{5,5',6,6'-tetrachloro-1-1'-diethyl-3,3'-di(4-sulfo-buthyl)-benzimidazolocarbocyanine} (TDBC) dyes. TDBC is the most widely used dye in plasmon-exciton coupling experiments due to its ability to organize in a J-aggregate fashion on gold surfaces \cite{DjoumessiLekeufack.2010, Zengin.2013}. The formation of these aggregates is beneficial for reaching the strong coupling regime, as J-aggregate excitons posses transition dipole moments far higher than the combined dipole moments of the constituting molecules.

\begin{figure*}
	\centering
	\includegraphics[width=1\textwidth]{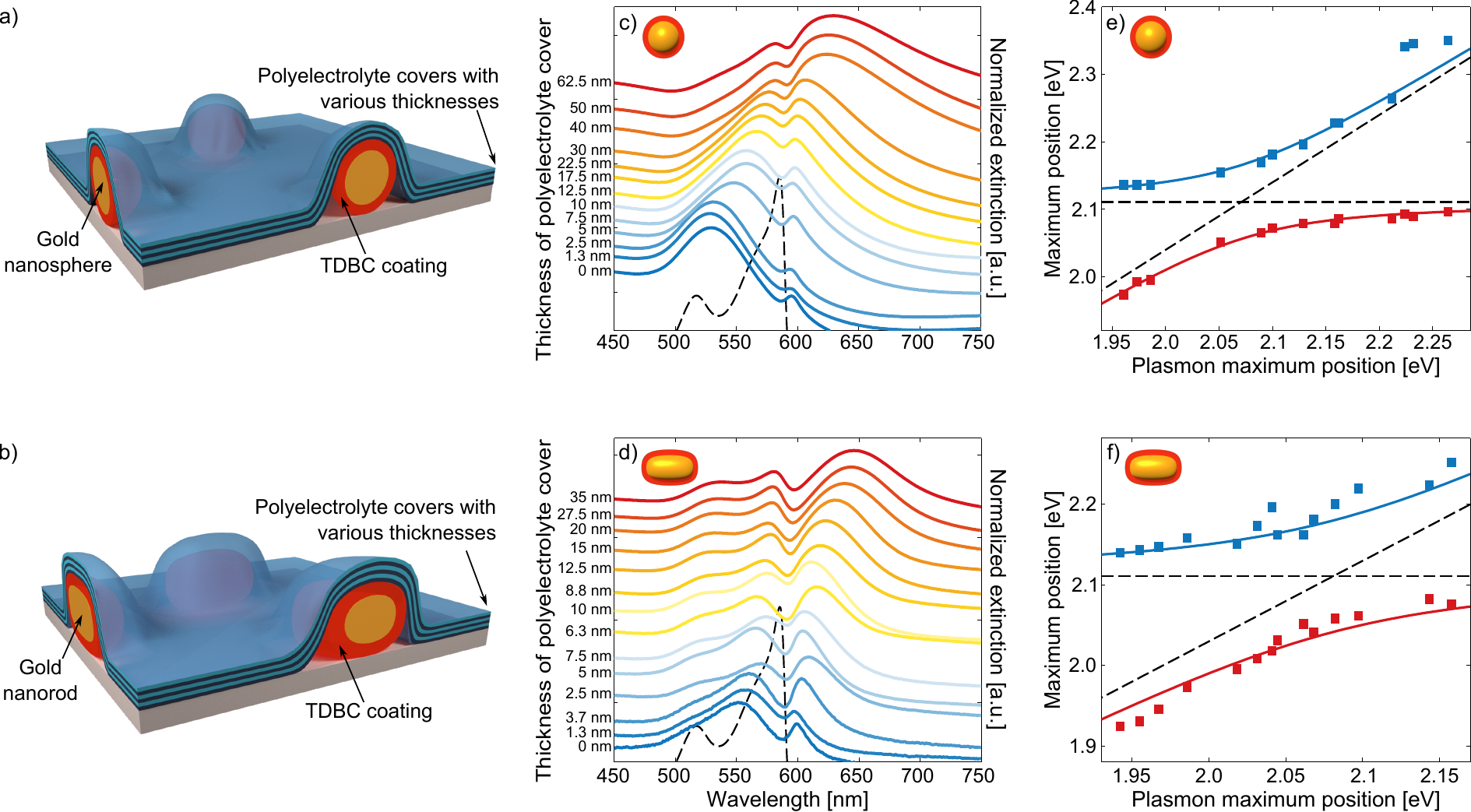}
	\label{fgr:TDBCAuNSandTDBCAuNRextspectraandanticrossing}
	\caption{Sketches of the samples of polyelectrolyte covered core-shell particles with a TDBC shell and a gold nanosphere (a) or nanorod (b) core. The extinction spectra for different cover thicknesses are presented in (c) for the spheres and in (d) for the rods together with the extinction of unaffected TDBC J-aggregates (black dashed line). A clear dip at the resonance position of the J-aggregates as well as a red-shift for thicker covers is visible. Maximum positions are plotted against the maximum positions of photo-bleached samples to reveal an anticrossing for both spheres (e) and rods (f). The red and blue lines are fits according to eq.1 with a small shift along the horizontal axis (see text.)}
\end{figure*}

Equation \ref{anticrossingequation} states that the resonance wavelengths of the coupled system strongly depend on the exciton-plasmon detuning $\delta = \omega_{p}-\omega_{ex}$. The essential parameter for discussing the coupling regime, is the resonance splitting $\Omega = 2\cdot g$, which describes the peak-splitting at $\delta = 0$. As assuring the spectral coincidence of exciton and plasmon resonance is difficult, the coupling regime is frequently discussed by plotting the resonances of the coupled system $\omega_{\pm}$ as a function of $\omega_{p}$. This procedure is similar to the measurement of the dispersion relation for coupled propagating plasmons and excitons, where the observation of an avoided crossing (anticrossing) of the plasmon and exciton dispersion relations is generally accepted as proof for strong coupling \cite{Novotny.2010,Torma.2015}. Conversely, observing an anticrossing of $\omega_{\pm}$ as a function of $\omega_{p}$ can be interpret as a sign for strong coupling in nanoparticle systems \cite{Fofang.2008, Ni.2008, Cade.2009, DeLacy.2013, Balci.2014, Zengin.2013, Melnikau.2016}.

The conventional way for tuning $\omega_{p}$ is changing the particle size \cite{Melnikau.2016} or, in case of nanoshells, the shell thickness \cite{Fofang.2008}. Using this approach to discuss the coupling strength has however several drawbacks. On the one hand, particles of only a few different particle sizes are usually available and hence the curves obtained this way consist only of a few points \cite{Melnikau.2016}. More importantly, different particle sizes are predicted to support different maximum coupling strengths \cite{Antosiewicz.2014}. Therefore, we consider the approach to change the particle size for discussing the coupling regime to be questionable. 

In contrast, our experimental approach consists of shifting the plasmon resonance by adjusting the permittivity $\epsilon_{med}$ of the particles' environment. The plasmon resonance is determined by the (dipole) polarizability $\alpha$ of the particle in this environment: \cite{Mitzscherling.2015, Bohren.1998}

\begin{equation}
\alpha \propto \frac{\epsilon_{mat}-\epsilon_{med}}{\epsilon_{mat}+f\epsilon_{med}}
\label{plasmonresonance}
\end{equation}

Here $\epsilon_{mat}$ denotes the permittivity of the nanoparticle, while the geometrical factor $f$ takes into account the shape of the particle. The plasmon resonance occurs at the wavelength for which the denominator becomes minimal. To change $\epsilon_{med}$ we embedded the particles in a polyelectrolyte-air matrix (see Figures 1a and 1b). The particles were deposited on a polymer-covered glass substrate and subsequently covered using layer-by-layer deposition of polyelectrolytes. Due to the low thickness of about $\SI{1.25}{nm}$ for each layer, the effective $\epsilon_{med}$ experienced by the particle is the average of the permittivity of the polymer cover and of the adjacent air. The step-wise addition of thin polymer layers then leads to an increase of the effective $\epsilon_{med}$, which in turn shifts the plasmon resonance \cite{Kiel.2012, Mitzscherling.2015}. We fabricated a separate sample for each cover thickness. This method allows a very fine tuning of the exciton-plasmon overlap, much more precise and facile than the tuning by particle size variation.

For nanoparticle systems the coupled resonances have been investigated in the past by extinction, transmission or reflection measurements \cite{Fofang.2008, Fofang.2011, DjoumessiLekeufack.2010, Melnikau.2016, Uwada.2007,  Ni.2008, Cade.2009, DeLacy.2013, Balci.2014, Zengin.2013, Zengin.2015, Zengin.2016}. In order to contrast these measurements to an approach based on the real particle absorption, we initially determined the extinction spectra of our samples as a function of  $\omega_{p}$. Extinction $E$ is a measure for the fraction of a light beam not transmitted through a sample. It can be measured as $E = 1 - T$, where the transmission coefficient $T=I_t/I_i$ is the ratio of incident versus transmitted intensity. The extinction spectra recorded for both particle types exhibit a dip at the spectral position of the J-aggregate absorption (Figures 1c and 1d). The rod spectra show an additional shoulder on the blue spectral side originating from the transverse LSP resonance. However, as it does not have a significant spectral overlap with the exciton absorption, it is only weakly coupled to the exciton and of no interest for the following discussion. Upon covering the hybrid particles with polymer layers, the spectral weight of the coupled extinction spectra shift to longer wavelengths. This is a consequence of the changing $\omega_{p}$, which leads to shifts of $\omega_{\pm}$ according to eq \ref{anticrossingequation}. For each cover thickness, the plasmon resonance $\omega_{p,b}$ was measured after photobleaching the TDBC in the very same samples that were used to obtain the coupled spectra. Compared to measuring $\omega_{p}$ on separate reference samples, the advantage of this procedure is that inhomogeneities in the sample structure are reflected in the measurements of both $\omega_{\pm}$ and $\omega_{p,b}$ likewise. The maximum cover thickness was chosen such that the plasmon resonance clearly shifted across the exciton resonance. 

Figures 1e and 1f present the energies $\hbar \omega_{\pm}$ describing the maximum positions of the extinction peaks versus the resonance energy $\hbar \omega_{p,b}$ of the plasmon. This yields a characteristic anti-crossing curve typical of two strongly-coupled oscillators. The coupling frequency $\Omega=2\cdot g$ corresponds to minimum distance between the branches. The horizontal dashed line indicates the exciton energy $\hbar \omega_{ex}$, which is independent of the cover layer thickness and represents an asymptotic solution of eq \ref{anticrossingequation} for $\omega_{p}-\omega_{ex}>>\Omega$. The tilted dashed line describes the asymptote corresponding to $\omega_p$, which is slightly shifted with respect to $\omega_{p,b}$. This is because bleaching the TDBC induces a chemical change in the molecule that alters $\epsilon_{med}$. Taking into account this shift by assuming $\omega_p \approx \omega_{p,b} - \SI{30}{meV} $ for rods and $\omega_p \approx \omega_{p,b} - \SI{40}{meV} $ for spheres, eq \ref{anticrossingequation} for the undamped coupled oscillator yields the excellent fit to the data in Fig. 1e,f. Considering this data only, one might therefore conclude that both systems are strongly coupled. However, extinction spectra contain information about the light scattered and absorbed by the particles likewise \cite{Quinten.2011}. The same is true for transmission measurements. These combined spectra are difficult to interpret, if the scattering and absorption fractions differ considerably from each other, which is the case for hybrid particles in the intermediate coupling-regime between weak and strong coupling \cite{Antosiewicz.2014}. In the following, we therefore discuss the coupling regime on the basis of the disentangled absorption and scattering spectra of the particles.

Absorption and scattering spectra of nanoparticles can be obtained using a spectrometer with an integrating sphere, which captures the total light scattered by the particles. Two measurements are necessary for each spectrum as illustrated in Figure 2. To obtain the absorption spectrum, the sum of the forward scattered and directly transmitted light intensity $I_{FS}+I_{T}$ is captured in a first measurement and the backscattered and reflected light $I_{BS}+I_{R}$ in a second.  The absorption is then $A=1-(I_{FS}+I_{T}+I_{BS}+I_{R})/I_i$. Similarly, the scattering spectrum $S$ is determined by measuring the forward and backward scattered light fraction without the transmitted beam and without specular reflection: $S=(I_{FS}+I_{BS})/I_i$. We measured absorption and scattering-spectra for the same set of samples for which the extinction spectra were obtained.

\begin{figure}
	\centering
	\includegraphics[width=0.45\textwidth]{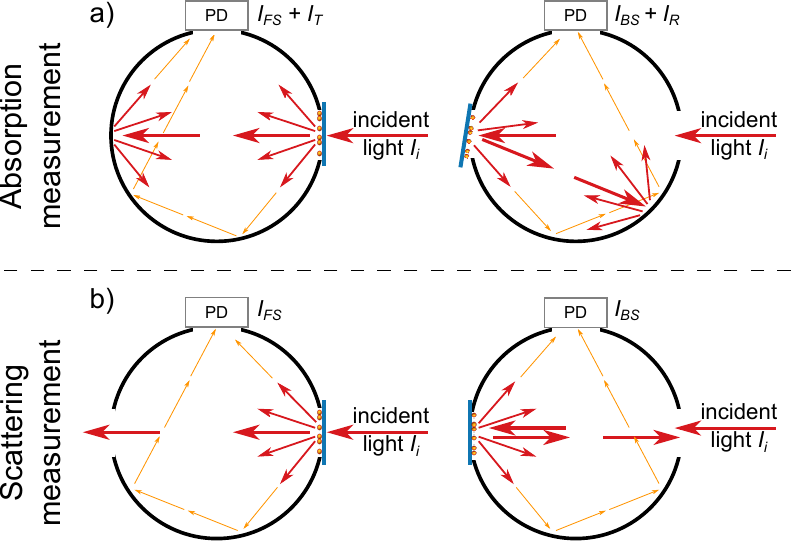}

	\caption{Configurations for measuring the absorption a) and scattering b) of a sample using a spectrometer with an integrating sphere.
\label{sch:Setup}}
\end{figure}

\begin{figure}
	\centering
	\includegraphics[width=0.49\textwidth]{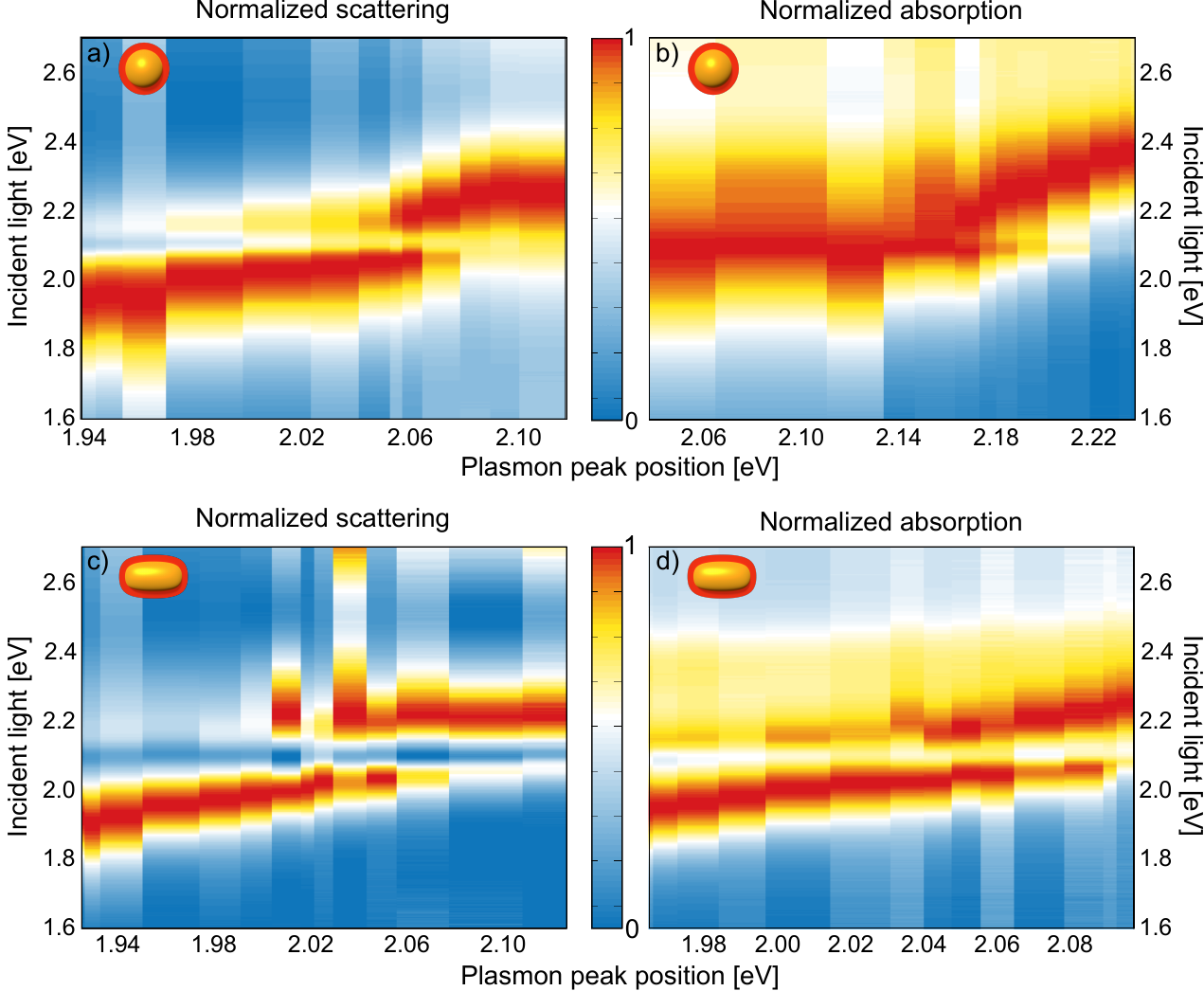}
	\caption{Scattering and absorption spectra of TDBC-coated gold nanospheres (top row) and gold nanorods (bottom row) covered with polyelectrolyte layers of various thickness. For the spheres only the scattering (a)) shows an anticrossing whereas in absorption (b)) the two branches do cross indicating that the system is not strongly coupled. For the rods both scattering (c)) and absorption (d)) show an anticrossing indicating that this system indeed is strongly coupled.
\label{fgr:scat+abs}}
\end{figure}

In Figure \ref{fgr:scat+abs}, we show the absorption $A(\hbar \omega_{pm})$ and scattering $S(\hbar \omega_{pm})$ spectra of TDBC-coated gold nanospheres and -rods with increasing polymer cover thicknesses in a 3D color plot, where the horizontal axis labels the increasing polymer cover thickness via the plasmon resonance position $\omega_{p,b}$. Only in the 3D plots of the rods' scattering spectra, we have subtracted contributions from clustered particles at near-infrared energies.

The scattering spectra for both particle types are comparable to the extinction spectra (spheres Figure \ref{fgr:scat+abs}a, rods Figure \ref{fgr:scat+abs}c). They show a spectral dip at the exciton resonance wavelength, while the spectral weights shift to lower energies for thicker polymer covers, caused by a shift of the plasmon resonance. The rods exhibit somewhat smaller linewidths and a deeper modulation than the spheres.
While the scattering spectra of both particles are qualitatively similar, the absorption spectra differ markedly. For the rods, also the absorption spectrum shows a behavior similar to the extinction, a dip and a corresponding anticrossing, as expected for a strongly coupled system. In contrast, the absorption spectrum for the spheres does not show a perceptible dip, but rather resembles a superposition of the separated exciton and plasmon absorption spectra, a behavior expected in the limit of weak coupling.

The reason for the differences between absorption and scattering spectra is rooted in the different optical cross-sections of the core compared to the shell. On the one hand, the immense scattering cross-sections of the plasmonic particle core vastly exceeds the cross-section of the adsorbed molecular aggregates \cite{Jain.2006}. Consequently, scattering from the core dominates the scattering-spectra of the hybrid particles, while scattering from the shell can be neglected for all practical purposes. The absorption spectrum, on the other hand, illustrates the energy dissipation in both core and shell of the particle. Even though the absorption cross-section of the core is much larger than the cross-section of the molecular aggregates, a large amount of energy is scattered from the plasmonic core to the shell. This process is well-known and is often employed to obtain single molecule spectra, e.g. by SERS \cite{Stiles.2008}. Thus a considerable fraction of the incoming light is transferred to and dissipated by the molecular shell. In other words, the experimental scattering spectrum only illustrates the behavior of the oscillator describing the plasmonic response, while the absorption spectrum includes the behavior of both plasmonic dissipation and dissipation in the TDBC shell \cite{Antosiewicz.2014}.

\begin{figure}
	\centering
	\includegraphics[width=0.49\textwidth]{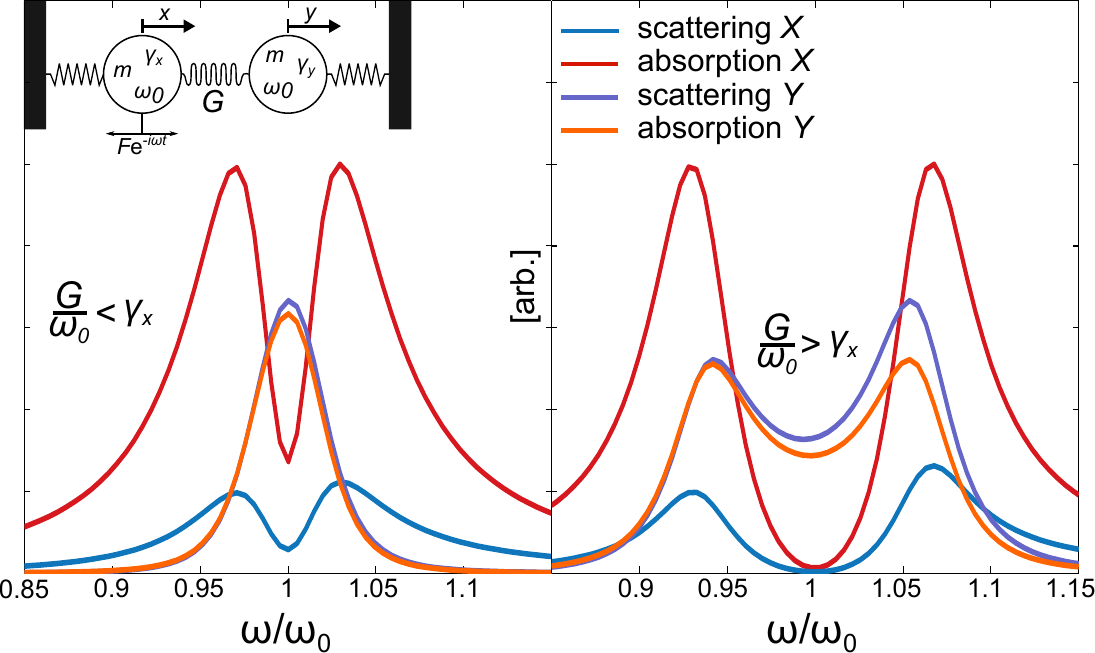}
	
	\caption{Absorption and scattering spectra of two classical coupled oscillators as shown in the inset on the top left for two different coupling strengths. Left: intermediate coupling with $G/\omega_0<\gamma_x$. Only the driven oscillator's resonance splits up. Right: strong coupling with $G/\omega_0>\gamma_x$. In this case both resonances show a splitting.
\label{fgr:oscisim}}
\end{figure}

In the following we will discuss this problem in a classical coupled oscillator model \cite{Wu.2010}. While the coupling of a single molecule to a single plasmon would require a quantum-mechanical description, the hybrid nanoparticles discussed here contain a large number of molecules and support a correspondingly large number of excitations. In this limit of many excitations the exciton-plasmon coupling is well described by a purely classical model  \cite{Torma.2015}. The coupled spring pendulum (Figure \ref{fgr:oscisim}, inset) presents the conceptually simplest mechanical equivalent to exciton-plasmon coupling in the limit of many excitations and is sufficient to discuss the origin of the splitting.

Two pendula X and Y, with resonance frequencies $\omega_i$, masses $m_i$ dampings $\gamma_i$, represent the core and shell resonances. The coupling is quantified by $G$ and is realized by a third spring between both oscillators. For simplicity, we discuss the case for which the oscillators are in resonance $\omega_X = \omega_Y = \omega_0$ and have the same masses $m_x=m_y=m$. The higher damping of the plasmon compared to the excitons is taken into account by a five times higher damping $\gamma_X$ for $X$, than the damping $\gamma_Y$ for $Y$. The system driven by an external force $F$ with a frequency $\omega$. Because of the higher optical cross-sections of the plasmonic nanoparticle core only the oscillator $X$ is excited by $F$. Mathematically, the coupled oscillators are described by their equations of motion:

\begin{equation}
\begin{split}
\ddot{x} + \gamma_{X} \dot{x} + \omega_0^2 x +G y &= \frac{F}{m} \mathrm{e}^{-i\omega t} \\
\ddot{y} + \gamma_{Y} \dot{y} + \omega_0^2 y +G x &= 0
\end{split}
\end{equation}

Here $x$ is the deflection of $X$ and $y$ the deflection of $Y$. The Fourier-Ansatz  $x(t) = X \mathrm{e}^{-i \omega t}$ and $y(t) = Y \mathrm{e}^{-i \omega t}$ and inversion of the resulting system of equations gives the corresponding complex amplitudes for both oscillators:

\begin{equation}
X =  \frac{\omega_0^2-\omega^2-i\gamma_{Y} \omega}{(\omega_0^2-\omega^2-i\gamma_{X} \omega)(\omega_0^2-\omega^2-i\gamma_{Y} \omega) -G^2}\frac{F}{m}
\end{equation}
and
\begin{equation}
Y =  \frac{G}{(\omega_0^2-\omega^2-i\gamma_{X} \omega)(\omega_0^2-\omega^2-i\gamma_{Y} \omega) - G^2}\frac{F}{m}
\end{equation}

To compare the pendulum model to the nanoparticle experiment, absorption and scattering have to be calculated from the complex oscillator amplitudes. The absorption $P_{\mathrm{loss}}$  for oscillator $X$ is given by the loss due to friction:

\begin{equation}
\overline{P_{\mathrm{loss}}} = \overline{ -m\gamma_X \dot{x}^2} = -\frac{m \gamma_X}{2} (\omega |X|)^2
\end{equation}

The power, $P_{\mathrm{scatt}}$, scattered by $X$, assuming the oscillators are of dipolar character is proportional to \cite{Bohren.1998}:

\begin{equation}
P_{\mathrm{scatt}} \propto \omega^4 |X|^2
\end{equation}
Identical formulas hold for oscillator $Y$.

Two criteria for the presence of strong coupling are regularly discussed in literature: $\Omega > \gamma_{Y},\gamma_{X}$ (strong criterion) and $\Omega > \sqrt{{\gamma_{Y}}{\gamma_{X}}}$ (weak criterion) \cite{Antosiewicz.2014, Faucheaux.2014, Cacciola.2014}. Taking into account that $\Omega = G/\omega_0$ \cite{Novotny.2010} Figure \ref{fgr:oscisim} illustrates the calculated absorption and scattering for $X$ and $Y$ in both regimes. If, on the one hand, the system fulfills the strong criterion (Figure \ref{fgr:oscisim}b), absorption and scattering of both oscillators undergo a splitting. This corresponds to the presence of new resonances for the coupled system representing new system eigenstates involving both oscillators, which is the most prominent characteristic of strong coupling. If, on the other hand, the system only fulfills the weak criterion, the amplitude spectrum of $X$ still splits, but the shape of $Y$ remains unmodulated (Figure \ref{fgr:oscisim} a). In this case, no system eigenstates can be defined and hence the system is not in the strong coupling regime \cite{Haroche.2006}. Antosiewicz et al. showed that a modulation of both oscillators is a necessary condition for strong coupling, and thus the system is only strongly coupled if it fulfills the strong criterion \cite{Antosiewicz.2014}. The weak criterion on the other hand represents an intermediate coupling regime, in which only the plasmon experiences a significant spectral modification.

The origin of the spectral splitting can be understood more intuitively by considering the excitation transfer in the time-domain. Let us consider the situation where $X$ is excited at its resonance frequency $\omega_0$: Due to the oscillator coupling, the motion of $X$ acts as an excitation force on $Y$. As the oscillators have the same (individual) resonance frequencies,  a phase-shift of  $\pi/2$ occurs between the oscillation of $X$ and $Y$. The coupling however works in both directions, hence the oscillator $X$ experiences a also feedback from $Y$. Since for $Y$ the coupling is stronger than all other decay channels, represented by $\gamma_Y$, the main part of the energy is transferred back to the first oscillator $X$. The respective periodic force exerted by oscillator $Y$ on oscillator $X$ again has a phase jump of $\pi/2$. Thus the total phase difference between the oscillation on $X$ induced by the external force and the feedback from oscillator $Y$ is exactly $\pi$. In other words, two out of phase oscillatory forces, which have opposite directions, act on oscillator $X$. This reduces the total oscillation amplitude of $X$ to values below those of the uncoupled oscillator, and possibly even to a complete suppression of the oscillation at this frequency. As a result a dip occurs in the amplitude spectrum of $X$.

In the discussion so far, there is no reason for $Y$ to split. Indeed, this interference on $X$ occurs already, if $\gamma_X \geq \Omega \geq \gamma_Y$. For $Y$ to show the same spectral behavior as $X$, a second feedback, this time from $X$ to $Y$, has to be possible. This means that the transfer from $X$ to $Y$ has to be faster than any other decay channel (or $\Omega > \gamma_X,\gamma_Y$). In this case, the two force components exerted from $X$ to $Y$ have a phase shift of $\pi$ and cancel each other out, such that no energy is transferred to $Y$ at all. From this discussion, we can conclude that for strong coupling to occur, at least one oscillation period relative to both oscillators has to be completed before the the dissipation essentially destroys the feedback.

In conclusion, we exemplified that absorption spectra reveal the true coupling regime of core-shell nanoparticles in a case where extinction spectra suggest wrong conclusions. We reported an approach for identifying the coupling strength by fine-tuning the resonances of hybrid exciton-plasmon particles via layer-by-layer deposition of polyelectrolytes. We selected two similar nanoparticle systems, TDBC coated hybrid nanospheres and -rods, which both exhibit extinction spectra with a pronounced dip and an anticrossing behavior. Careful distinction of transmission, reflection and scattering allows for measuring the pure absorption which revealed that only in the rod-like particles the plasmon resonance was strongly coupled to the excitons. In order to understand the physical mechanism for this behavior we discussed the analog of a classical coupled oscillator model, where only one oscillator is directly excited by the driving light field. The model clarifies that the coupling induces a feedback between the two oscillators, which only leads to a dip in the dissipation spectrum describing absorption, if the coupling is not only strong enough to transfer the energy form the driven oscillator to the "dark" oscillator. The energy must efficiently be transferred once more to the driven oscillator and back to the dark oscillator, before the phase information is lost by dissipation. In contrast, the scattering spectrum of hybrid particles is dominated by the plasmon contribution and the negative feedback already shows up for an intermediate coupling which transfers the energy to the dark oscillator and back once. In ambiguous situations, in which the splitting of the extinction is similar to or smaller than the linewidth of the unmodulated plasmon peak, only an absorption spectroscopy that accounts for the scattering appropriately, can conclusively distinguish between strong and intermediate coupling. We expect that these results will facilitate the further development of strongly-coupled plasmon-exciton hybrid nanoparticles by assisting the community to unambiguously distinguish between the strong and intermediate coupling regimes.

\vspace{5mm}

\textit{\textbf{Methods:}} TDBC was purchased from FEW-chemicals, the gold nanospheres (diam $\SI{100}{nm}$, ligand citrate), poly-allylamine hydrochloride, poly-sodium 4-styrenesulfonate, poly-ethyleneimine, and Tween-20 were purchased from Sigma-Aldrich, gold nanorods (res. 25-600, ligand citrate) came from Nanopartz.

\textit{{Coating of nanoparticles with TDBC}}: The fabrication process of TDBC coated nanoparticles mainly followed the approach by Lekeufack et al. \cite{DjoumessiLekeufack.2010}, however to prevent the formation of clusters an intermediate coating with a non-ionic surfactant was established \cite{Aslan.2002}: The TDBC was dissolved in aqueous NaOH solution ($c_{NaOH}=\SI[exponent-product=\cdot]{e-5}{mol/l}$) to obtain a concentration of ca. $c_{dye} \approx \SI{1}{mmol/l}$. The mixture was stirred for 5 minutes and placed in an ultrasonic bath for 15 minutes.

The gold nanoparticles had a negative surface charge. To prevent clustering with positively surface charged TDBC-coated particles, $\SI{1}{ml}$ of particle solution was mixed with $\SI{20}{\textmu l}$ of the non-ionic surfactant Tween-20 and left to rest for two hours. This mixture was then added to the TDBC solution. The ratio between particle solution and TDBC solution was 1:1. After an ultrasonic bath with a duration of 7 minutes the mixture was stored for 48 hours.

After the resting time the solutions were centrifuged twice at $\SI{3000}{rpm}$ for 30 minutes (rods) or $\SI{4000}{rpm}$ for 20 minutes (spheres). After the second centrifugation and removal of the excess particles were redissolved in water: rods in 0.7 times the excess volume, spheres in 0.25 times the excess volume.

\textit{{Deposition of particles on substrate}}: For functionalisation of substrates and the creation of polyelectrolyte layers poly-ethyleneimine (PEI,cationic), poly-sodium 4-styrenesulfonate (PSS,anionic) and poly-allylamine hydrochloride (PAH,cationic) were used. The polymers were dissolved in a NaCl solution with a polymer concentration of $\SI{1}{wt\%}$ (PEI) or $\SI{0.1}{wt\%}$ (PSS,PAH) and a salt concentration of $\SI{0.7}{mol/l}$. Glass substrates were cleaned in ultrasonic methanol bath for 15 minutes and subsequently washed in ultrasonic water bath for 15 minutes. They were then funcionalized by spin coating one layer of PEI, followed by a layer of PSS. To create a homogeneous monolayer 5-7 drops of a polymer solution were deposited on the substrate spinning at a rotation speed of $\SI{3000}{rpm}$ and after a few seconds washed way with 5-7 drops of water. Subsequently $\SI{350}{\textmu l}$ of TDBC-nanoparticles were deposited on the substrate. After an adsorption time 4 hours for spheres, 12 hours for rods) excess particles were rinsed away with water and the desired amount of alternating layers of PSS and PAH (starting with PSS) was spin coated on top.

\textit{Photobleaching of TDBC}:  TDBC was photobleached with a cw-laser working at $\SI{532}{nm}$ at $\SI{10}{W}$. The laser beam was widened to an area an area of approximately $\SI{1}{cm^2}$ to photobleach one whole sample simultaneously and not melt the gold particles. The bleaching time was 8 hours.

\textit{{Measurements}}: Absorption and scattering spectra were recorded in a \textit{Cary 5000} spectrometer, extinction measurements were executed in a \textit{Cary 5e} spectrometer.

\begin{acknowledgements}
FS acknowledges financial support by the DFG via the graduate school SALSA.
\end{acknowledgements}

\bibliography{references}

\end{document}